\documentclass[aps,twocolumn,prb]{revtex4}     
\usepackage{graphicx}                   

\begin{document}

\title{A combined theoretical and experimental study of the low temperature  properties of BaZrO$_3$ }

\author{A. R. Akbarzadeh$^{1}$,  I. Kornev$^{1}$, C. Malibert$^{2}$,
L. Bellaiche$^{1}$, and J. M. Kiat$^{3}$}

\address{$^{1}$Physics Department, University of Arkansas, Fayetteville, AR 72701, Arkansas, USA}
\address{$^{2}$Laboratoire d'Etudes des Milieux Nanom$\acute{e}$triques,
Universit$\acute{e}$d'Evry Val d'Essonne, 91000 Evry, France}
\address{$^{3}$ Laboratoire L$\acute{e}$on Brillouin, CE Saclay, 91191
Gif-sur-Yvette Cedex, France}
\date{\today}

\begin{abstract}

Low temperature properties of BaZrO$_3$ are revealed by combining
experimental techniques (X-ray diffraction, neutron scattering and
dielectric measurements) with theoretical first-principles-based
methods (total energy and linear response calculations within
density functional theory, and effective Hamiltonian approaches
incorporating/neglecting zero-point phonon vibrations). Unlike
most of the perovskite systems, BaZrO$_3$ does not undergo any
(long-range-order) structural phase transition and thus remains
cubic and paraelectric down to 2\,K, even when neglecting
zero-point phonon vibrations. On the other hand, these latter pure
quantum effects lead to a negligible thermal dependency of the
cubic lattice parameter below $\simeq$ 40\,K. They also affect the
dielectricity of BaZrO$_3$ by inducing an overall saturation of
the real part of the dielectric response, for temperatures below
$\simeq 40\,$K. Two fine structures in the real part, as well as
in the imaginary part, of dielectric response are further observed
around 50-65\,K and 15\,K, respectively. Microscopic origins
(e.g., unavoidable defects and oxygen octahedra rotation occurring
at a local scale) of such anomalies are suggested. Finally,
possible reasons for the facts that some of these dielectric
anomalies have not been previously reported in the better studied
KTaO$_3$ and SrTiO$_3$ incipient ferroelectrics are also
discussed.
\end{abstract}



\narrowtext

\maketitle

\section{INTRODUCTION}
\label{s:intro}

BaZrO$_3$ is a ceramic oxide of the perovskite family structure
with a large lattice constant, high melting point, small thermal
expansion coefficient, low dielectric loss and low thermal
conductivity (see, e.g.
Refs.~\onlinecite{BhallaAPL,Majeed1,Majeed2,Robertz,Lecerf,Taglieri,
Celik,Koopman,Brzezinska,ChaiDavies,Erb,Kiyotaka,Davies} and
references therein). These afore-mentioned properties make
BaZrO$_3$ (i) a very good candidate to be used as  an inert
crucible in crystal growth techniques \cite{Celik,Erb}, (ii) an
excellent material for wireless
communications~\cite{BhallaAPL,ChaiDavies} and (iii) a very good
substrate in thin film deposition~\cite{BhallaAPL,Majeed2}.
BaZrO$_3$ is also one of the two parent compounds of the (Pb-free
and thus environmental-friendly)  Ba(Zr,Ti)O$_3$ solid solutions,
which is promising for manufacturing high Q materials with a
variety of applications in microwave industry~\cite{Davies}.

Interestingly, properties of BaZrO$_3$ have been measured as long
as 40 years ago, as well as very recently
~\cite{BhallaAPL,Majeed1,Majeed2}, but only at room or higher
temperature (to the best of our knowledge). Similarly, we are not
aware of any calculation (either from phenomenological theory or
first-principles calculations) predicting the dielectric
properties of BaZrO$_3$. As a result, low-temperature dielectric
properties of BaZrO$_3$ have never been investigated, despite the
fact that many unusual effects are known to occur in  some
perovskite materials between 0 and 50\,K. One drastic example of
such effects is the (temperature-independent) plateau and large
values of the real part of the dielectric response in KTaO$_3$ and
SrTiO$_3$, which arise from the quantum-induced suppression of
ferroelectricity in these materials. Other examples are the
anomalous peaks observed around 10-50\,K for the imaginary part of
the dielectric response in KTaO$_3$, K(Ta,Nb)O$_3$,
(Pb,La)TiO$_3$:Cu~\cite{Salce,Bidault},
SrTiO$_3$~\cite{Bidault,Viana}, that are neither associated with
structural phase transition nor do have a corresponding peak in
the real part of the dielectric response (which conflicts with the
well-established Kramers-Kronig relations~\cite{Scaife}).

The aim of this article is to investigate the low-temperature
behavior of BaZrO$_3$ from measurements and first-principles-based
simulations. We report several unusual features in the real and
imaginary parts of the dielectric responses while {\it no}
long-range ferroelectric, antiferroelectric or antiferrodistortive
structural phase transition occurs in BaZrO$_3$ down to 2\,K.
Discussions and similarities/differences between BaZrO$_3$ and the
(better-studied) KTaO$_3$ and SrTiO$_3$-related materials are also
indicated to better understand the low-temperature dielectric
anomalies reported in several perovskites.

This article is organized as follows. Sec.~\ref{s:method}
describes the experimental and theoretical methods we used to
investigate BaZrO$_3$. Sec.~\ref{s:Res} reports the measurements
and predictions for structural and dielectric properties.
Finally, Sections ~\ref{s:disc} and
~\ref{s:concl} provide a discussion and conclusion, respectively.

\begin{table*}
\caption{The LDA-derived H$_{\rm eff}$ parameters in atomic units
for BaZrO$_3$ following the notation in Ref~[29].}
\label{t:THeffBZ}
\begin{tabular}{cccccccc}
  \hline \hline
lattice constant &  & a$_0$ & 7.91943 & Soft mode mass & 75.721 & & \\
\hline
Onsite & & $\kappa_2$ & 0.0183 & $\alpha$ & 0.009733 & $\gamma$ & 0.01663 \\
\hline
& & $j_1$ & 0.00738 & $j_2$ & 0.02311 & &\\
Intersite & &$j_3$ & 0.00262 & $j_4$ & -0.00163 & $j_5$ & 0.00120 \\
& &$j_6$ & 0.00049 & $j_7$ & 0.00024 & & \\
\hline
Elastic & & $B_{11}$ & 4.794 & $B_{12}$ & 0.755 & $B_{44}$ & 1.416 \\
\hline
Soft mode-elastic & & $B_{1xx}$ & -0.431 & $B_{1yy}$ & 0.033 & $B_{4yz}$ & -0.055 \\
 \hline
Dipole & &  $Z^*$ & 5.81 & $\varepsilon_\infty$ & 4.928\\
 \hline \hline
\end{tabular}
\end{table*}
\section{METHODOLOGY}
\label{s:method}

\subsection{EXPERIMENTAL PROCEDURES}
\label{s:exp}

Powder samples of BaZrO$_3$ were synthesized by solid state
reaction by calcination at 1400K and sintering at 1600K starting
from stoichiometric amounts of the corresponding oxides (BaCO$_3$,
ZrO$_2$). The synthesized samples were well crystallized and no
presence of parasitic phases was evidenced by X-ray diffraction
(XRD) and chemical analysis. The temperature dependence of the
dielectric constant was measured at various frequencies in a
temperature range from 5\,K to 300\,K using a Hewlett-Packard
4192A impedance analyzer and a cryostat with an estimated
precision of 0.1 K. These measurements were performed on ceramic
samples which were polished and cleaned, and sputtered gold
electrodes were applied.  Samples were annealed at 800\,K and
slowly cooled in order to eliminate strains caused by polishing.
Powdered samples were used for the diffraction experiments.
X-ray-diffraction measurements were performed on a high accuracy,
two-axis diffractometer in a Bragg-Brentano geometry using Cu-Kb
wavelength issued from a 18-kW rotating anode generator, with
diffraction angles precision better than 0.002 $\deg$. The neutron
powder diffraction patterns were collected at temperatures between
300 and 2\,K on the 3T2 high resolution goniometer on a thermal
source (1.227 $\AA$) using the OrphŽe reactor facilities at
Laboratoire L\'eon Brillouin (Saclay, France). Structural Rietveld
refinements on both X-ray and neutron patterns were carried out
with the XND software.

\subsection{THEORETICAL APPROACHES}
\label{s:theory}

In this study, two different (direct) first-principles codes, as
well as a first-principles-derived technique,  were used to obtain
and/or extract various information. One of the two
first-principles codes is denoted ABINIT \cite{ABINIT}. We took
advantage of its implementation of the linear response theory to
compute the phonon dispersion curves of BaZrO$_3$ at its
low-temperature experimental lattice  constant. Some technical
details are as following. In this method we used Teter extended
norm-conserving pseudopotentials~\cite{Teter} and the
local-density approximation (LDA) \cite{Kohn}. The
exchange-correlation functional was approximated using
Perdew-Zunger parametrization \cite{PZ} of Ceperley-Alder data
\cite{CA}. The Ba $5s^2$, Ba $5p^6$, Ba $6s^2$, Zr $4s^2$, Zr
$4p^6$, Zr $4d^2$, Zr $5s^2$, O $2s^2$, and O $2p^6$ electrons are
treated as valence electrons. A plane-wave cut off of 100\,Ry was
used to obtain convergence of the ground state total energy.
Phonon frequencies were found to converge for dynamical matrices
calculated on a 6$\times$6$\times$6 Monkhorst-Pack grid \cite{MP}.

The second first-principles code used in our study is denoted CUSP
\cite{CUSP}. It also implements the LDA \cite{Kohn}, and the
Ceperley-Alder exchange and correlation \cite{CA} as parameterized
by Perdew and Zunger \cite{PZ}. We also used the same valence
electrons (indicated above) as in the ABINIT code. On the other
hand, the pseudopotentials are those given by the Vanderbilt
ultrasoft scheme \cite{USPP}, and the plane-wave cutoff is chosen
to be 25\,Ry, which leads to converged results of physical
properties of interest \cite{CUSP}. The CUSP code is used to
calculate the energetics of BaZrO$_3$ that are associated with the
rotation of the oxygen octahedra,  at its low-temperature
experimental lattice constant. It is also used to derive, at this
specific lattice constant, the 18 parameters of the
first-principles-based effective Hamiltonian ($H_{\rm eff}$)
approach developed in Ref.~\onlinecite{ZhongDavid}. These
parameters are given in Table I.

We also perform simulations using such $H_{\rm eff}$ technique
(and its parameters) to go beyond the abilities of direct
first-principles techniques, namely to investigate {\it
finite-temperature} properties of large BaZrO$_3$ supercells. The degrees of
freedom of this Hamiltonian are the so-called local modes (that
are directly proportional to the spontaneous electrical polarization), and
the strain variables (that characterize the crystallographic
phase). The total energy, E$_{tot}$ of H$_{\rm eff}$ contains five
different interactions between local modes and/or  strains,
namely, a local-mode self energy, a long-range dipole-dipole
interaction, a short-range interaction between local modes, an
elastic energy and an interaction between local-modes and strain
~\cite{ZhongDavid}. Monte-Carlo (MC) simulations are performed
using $E_{tot}$ in two different schemes: classical Monte Carlo
(CMC)~\cite{Metropolis}, which neglects zero-point phonon
vibrations and, path-integral quantum Monte Carlo (PI-QMC)
~\cite{jorge,ZhongPRB96,ceperley}, which includes these
quantum-mechanical zero-point motions. Consequently, comparison
between the results of these two different Monte-Carlo techniques
will allow the determination of quantum effects on structural
and dielectric properties of BaZrO$_3$.
12$\times$12$\times$12 supercells (corresponding to
8,640 atoms) are used in all Monte-Carlo simulations. We typically
perform 30,000 MC sweeps to thermalize the system and 70,000 more
to compute averages, except at low temperatures in PI-QMC where
more statistics is needed.

In PI-QMC, each 5-atom cell interacts with its images at neighboring
imaginary times through a spring-like potential (representing the
zero-point phonon vibrations as implemented in PI-QMC formulations),
while all the 5-atom cells interact with each other at
 the same imaginary time through the internal potential associated with $E_{\rm tot}$.
The product $TP$, where $T$ is the simulated temperature and $P$
is the number of imaginary time slices, controls the accuracy of
the PI-QMC calculation~\cite{Cuccoli}. In all our simulations we
use $TP$=600, which we find leads to sufficiently converged
results. Outputs of the PI-QMC simulations thus contain
local-modes {\bf u$_i$}(t), where $i$ runs over the 5-atom unit cells of
the studied supercell while the imaginary time t ranges between 1
and $P$. (Note that t $=P=1$ corresponds to CMC simulations.)
Strain variables are another output of MC simulations.

\section{RESULTS}
\label{s:Res}
\subsection{STRUCTURAL PROPERTIES}
\label{s:Resstruc}
XRD and neutron scattering indicate that BaZrO$_3$ is cubic (and
thus paraelectric) down to 2 Kelvin (the minimal temperature
accessed during our neutron scattering). Note that we are not
aware of any previous measurement investigating the
low-temperature, rather than high-temperature~\cite{Majeed1,Majeed2,Fuenzalida},
properties of BaZrO$_3$.

\begin{figure}
\includegraphics[height=0.3\textheight]{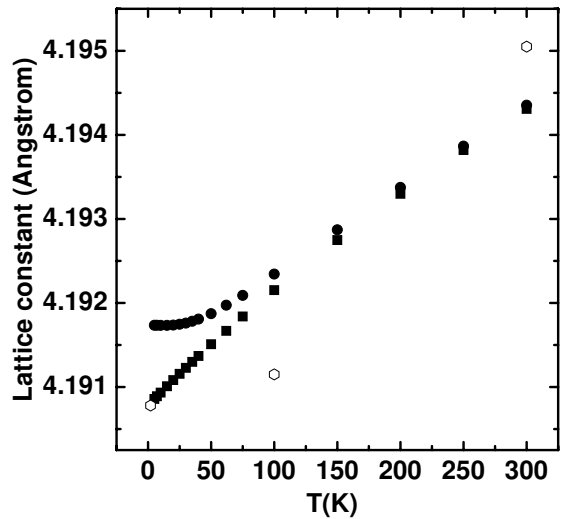}
\caption{Evolution of the cubic lattice constant as a function of
temperature in BaZrO$_3$. Solid squares and solid circles
correspond to the CMC and PI-QMC predictions, respectively, while
the open symbols display the results of neutron scattering. Note
that the CMC predictions exactly agree with the experimental data
at low temperature as a consequence of our choice of the a$_0$
parameter for the effective Hamiltonian approach (see Table
I).}\label{f:AlatT}
\end{figure}

Our experimental result is consistent with the fact that we
further find that our PI-QMC simulations using the effective
Hamiltonian approach provide a vanishing supercell average of the
local-modes down to 0\,K and a resulting cubic ground-state.
Interestingly, our CMC simulations also predict such paraelectric
ground-state. This indicates that, unlike
KTaO$_3$~\cite{AliLaurent,samara} and SrTiO$_3$
\cite{ZhongPRB96,Viana} for which quantum effects suppress
ferroelectricity in favor of paraelectricity, zero-point phonon
vibrations do {\it not} affect the symmetry of the ground-state in
BaZrO$_3$. On the other hand, Fig.~\ref{f:AlatT} -- which displays
the predicted and measured temperature evolutions of the cubic
lattice constant of BaZrO$_3$ -- clearly shows, via the comparison
of CMC and PI-QMC results, how zero-point phonon vibrations {\it
quantitatively} affect structural properties of BaZrO$_3$. Below
100\,K, the zero-point phonon vibrations tend to prevent the
lattice constant from decreasing when decreasing the temperature.
Quantum effects thus increase the lattice constant with respect to
classical predictions, with this increase becoming more pronounced
as the temperature decreases. As a matter of fact, PI-QMC
calculations result in (1) a dramatic change of thermal expansion
below {\it versus} above 100\,K, (2) a lattice constant  that is
nearly temperature-independent below $\simeq$ 40 \,K, and (3) a
lattice parameter that is larger by $10^{-3}$ $\AA$ from its CMC
result at the lowest temperatures. Note that item (1) is clearly
confirmed by our neutron scattering data (reported in
Fig.~\ref{f:AlatT} for 300, 100 and 2\,K) that shows a relatively
rapid $\simeq 10^{-3} \AA$ decrease of the lattice constant
between 300\,K and 100 \,K, while the difference in lattice
parameters between 100 and 2\,K is as small as $3 \times 10^{-4}
\AA$. Furthermore, items (1) and (2) have also previously been
observed in the incipient KTaO$_3$ system~\cite{SM} (but for
different critical temperatures). Moreover, item (3) provides a
measure of the (relatively small) quantitative effects of
zero-point phonon vibrations on the lattice parameter {\it per se}
(rather than on the thermal expansion) of  BaZrO$_3$.

Fig.~\ref{f:AlatT} also indicates that above 100 \,K, one can
safely use the approximation that the cubic lattice parameter {\it
linearly} depends on temperature with the thermal expansion
coefficient being 0.27 $\times 10^{-5}$ (1/K), 0.23 $\times
10^{-5}$(1/K) and 0.47 $\times 10^{-5}$(1/K) for the CMC, PI-QMC
and neutron scattering data, respectively. Our experimental value
of 0.47 $\times 10^{-5}$(1/K) compares rather well with the linear
thermal expansion coefficient of 0.69 $\times 10^{-5}$(1/K)
previously measured by X-ray diffraction for temperature ranging
between 273 and 873\,K ~\cite{Zhao}. On the other hand, the
thermal expansion of  0.27 $\times 10^{-5}$ (1/K) and 0.23 $\times
10^{-5}$(1/K) predicted by our simulations underestimate the
experimental values by a ratio of $2-3$. This discrepancy between
simulations and the measurements  arises from the fact that the
effective Hamiltonian approach only incorporates the
ferroelectric-related vibrations among the optical modes, while an
accurate description of thermal expansion requires to take into
account {\it all} phonon modes \cite{Tinte}.

\begin{figure}
\includegraphics[height=0.3\textheight]{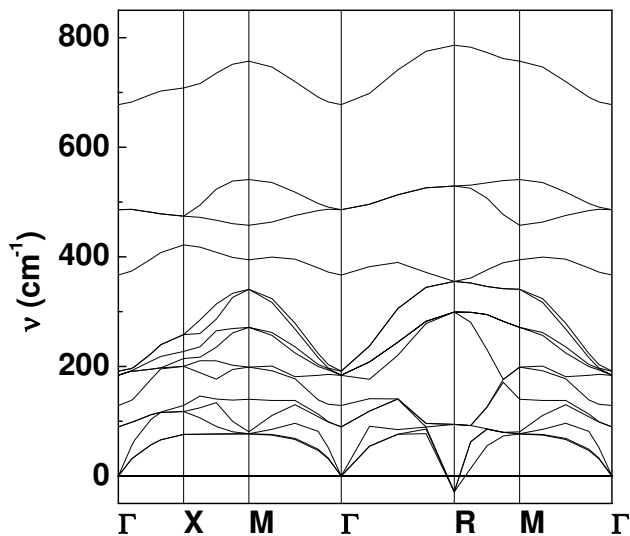}
\caption{ Phonon dispersion of cubic BaZrO$_3$, from first
principles calculations.}\label{f:fig1BZ}
\end{figure}

\begin{figure}
\includegraphics[height=0.3\textheight]{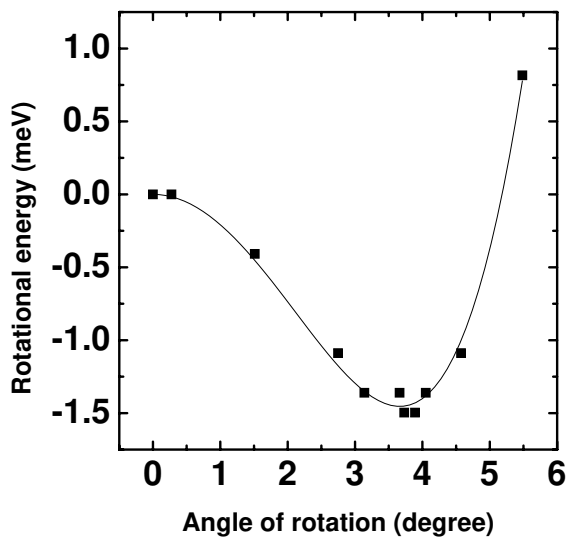}
\caption{ Oxygen octahedra rotational energetics (per 5 atoms)
\emph{versus} the angle of the rotation about the [001] direction
from first principles calculations. The zero energy correspond to
the paraelectric state. Solid line is a guide for the
eyes.}\label{f:fig2BZ}
\end{figure}

The effective Hamiltonian approach used in our study thus also
neglects some phonon modes that may condense in BaZrO$_3$, such as
the R$_{25}$ modes that are associated with the rotation of the
oxygen octahedra ~\cite{ZhongDavidPRL}. To check such possibility,
we decided to compute the whole phonon dispersion of cubic
BaZrO$_3$, using the linear response theory as implemented in the
first principles ABINIT code \cite{ABINIT}. The results are shown
in Fig.~\ref{f:fig1BZ}, and indeed confirm the previous {\it
ab-initio}  prediction \cite{ZhongDavidPRL} of the condensation of
the R-point zone-boundary mode.
(Note that these phonon calculations also confirm that BaZrO$_3$
does not exhibit any ferroelectric instability at the $\Gamma$
point, even when quantum statistics are neglected). One has to
realize, though, that the instability associated with these
antiferrodistortive motions is rather weak, as demonstrated by the
relatively {\it small} negative value of the frequency. This is
further evidenced in Fig.~\ref{f:fig2BZ}, that displays the total
energy (as predicted by the first principles CUSP program
\cite{CUSP}) versus the angle of the octahedra rotation about the
[001] direction: the minimum energy (occurring for an angle about
4 $\deg$) is only $\simeq$ 1.5 \,meV deeper than the energy of the
paraelectric phase (associated with a zero angle). Such difference
in energy corresponds to a rather small temperature of
$\approx$17\,K. It is thus highly possible that zero-point phonon
vibrations prevent the occurrence (down to 2\,K) of the
macroscopic cubic paraelectric--to--antiferrodistortive phase
transition, which would explain why our low-temperature XRD and
neutron scattering experiments do not reveal any additional peak
related to a doubling of the unit cell. This quantum-induced
suppression may act either globally (i.e., different unit cells do
exhibit some rotation of their oxygen octahedra, but these
rotations are not long-range correlated) or locally (i.e., there
is no rotation of the oxygen octahedra in any unit cell). We will
come back to this point, and to this $\approx$17\,K temperature,
when discussing the results on dielectric properties.

\subsection{REAL PART OF THE DIELECTRIC RESPONSE}
\label{s:Resreal}

\begin{figure}
\includegraphics[height=0.3\textheight]{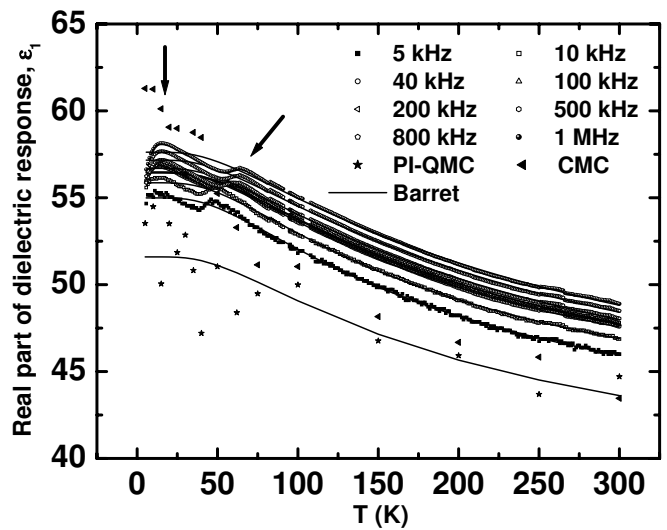}
\caption{ Theoretical (using CMC and PI-QMC simulations) and
experimental (for different frequencies) real part of the
dielectric constant. Solid lines show the fit of the data into the
Barret relation. The arrows indicate the position of the two
observed dielectric anomalies that are discussed in the
text}\label{f:fig3BZ}
\end{figure}

Fig.~\ref{f:fig3BZ} shows our experimental determination of the
real part of the dielectric response, $\varepsilon_1$,
\textit{versus} temperature for different frequencies.
\begin{table*}
\caption{Coefficients of the fit of the real part of the
dielectric response into the Barrett relation. All the parameters
of all the fits are allowed to relax, except the parameters C and
T$_s$ for the fit of the PI-QMC data that are kept frozen to their
deduced values from our experiments at the lowest frequency. Such
freezing is necessary to allow the fit to converge because of the
relatively large fluctuation inherent to PI-QMC simulations
 at low temperature.}\label{t:TBarrett}
\begin{tabular}{llllllllll}
\hline\hline
 f[kHz] & 5 & 10 & 40 & 100 & 200 & 500 & 800 & 1000 & PI-QMC \\
 \hline
 {\it C} & 2900$\pm$185 & 2900 $\pm$170 & 3200$\pm$200 & 3400$\pm$240 &
 3600$\pm$280 & 4200$\pm$370 & 4600$\pm$440 & 4800$\pm$480 & 2900 \\
  {\it T$_s$}[K] & 81$\pm$2 & 80$\pm$2 & 80$\pm$2 & 80$\pm$2 &
 80$\pm$2 & 77$\pm$3 & 76$\pm$3 & 75$\pm$3 & 81  \\
  {\it T$_0$}[K] & -99$\pm$10 & -102$\pm$9 & -114$\pm$10 & -124$\pm$12 &
 -135$\pm$13 & -160$\pm$17 & -175$\pm$19 & -182$\pm$20 & -114$\pm$21 \\
 {\it B} & 38.8$\pm$0.3 & 39.6$\pm$0.2 & 40.0$\pm$0.3 & 39.8$\pm$0.4 &
 39.5$\pm$0.4 & 39.0$\pm$0.5 & 39.0$\pm$0.6 & 39.0$\pm$0.6 & 36.7Ê$\pm$ 1.4
\\
 \hline\hline
 \end{tabular}
\end{table*}
The real part of the dielectric response of BaZrO$_3$ increases as
the temperature decreases from 300\,K for any given frequency. It
then exhibits an overall low-temperature saturation to a plateau
having a value that is slightly dependent on the frequency
(ranging between 55 and 58 for frequency varying between 5\,kHZ
and 1\,MHz). The existence of such plateau has already been
reported in the incipient ferroelectrics
KTaO$_3$~\cite{AliLaurent,samara} and
SrTiO$_3$~\cite{ZhongPRB96,Viana}, and is usually thought to be
associated with zero-point phonon vibrations. To check this fact,
we also report in Fig.~\ref{f:fig3BZ} the predictions from the
$H_{\rm eff}$ approach using both CMC and PI-QMC techniques. It is
obvious that using CMC, the real part of the dielectric response
continuously increases to higher value when the temperature
decreases while on the other hand the dielectric constant computed
within PI-QMC tends to saturate, in overall, at
low-temperature~\cite{footnote}. Our simulations thus prove that
quantum fluctuations play an important role on the low-temperature
dielectric responses of BaZrO$_3$, as in
KTaO$_3$~\cite{AliLaurent,samara} and
SrTiO$_3$~\cite{ZhongPRB96,Viana}. However and as already
mentioned in Section III.A, this quantum-induced modification of
dielectricity in BaZrO$_3$ is {\it not} accompanied by a
suppression of ferroelectricity since BaZrO$_3$ does not have any
ferroelectric instability even in the classical regime -- as
indicated by the fact that our CMC simulations lead to a cubic
paraelectric ground-state and a lack of peak in $\varepsilon_1$.
This distinguishes BaZrO$_3$ from both KTaO$_3$ and SrTiO$_3$.
Figure~\ref{f:fig3BZ} also reveals that our PI-QMC simulations
yield predictions that are slightly smaller in magnitude than our
experimental results. In fact, this has to be expected since (i)
the measurements displayed in Fig.~\ref{f:fig3BZ} demonstrate that
decreasing frequency leads to a decrease of the dielectric
response at any temperature and (ii) the calculations correspond
to the static regime, \textit{i.e.} to a zero frequency. We can
thus conclude that the PI-QMC results are in overall quite
accurate, especially when realizing that dielectric coefficients
are related to the {\it derivative} of the polarization (i.e.,
they are much more difficult to predict than properties that are
directly proportional to polarization) and that the magnitude of
$\varepsilon_1$ is rather small in BaZrO$_3$ (e.g., the
low-temperature plateau has a value that is smaller by $\simeq$
2-3 orders of magnitude than the corresponding ones in KTaO$_3$
~\cite{AliLaurent,Salce,SamaraBook}and SrTiO$3$
~\cite{KleemannQBarret,Viana})

To further analyze our results, we
fitted them using the Barret relation~\cite{Barrett}
\begin{eqnarray}~\label{eq:CBarret}
\varepsilon_1=\frac{C}{T_s coth(\frac {T_s}{T})- T_0} + B
\end{eqnarray}
where $C$ and $T_s$ are the so-called Curie constant and
saturation temperature, respectively
~\cite{Barrett,AliLaurent,Ang}. $T_0$ is interpreted as being the
classical Curie temperature (that is, the Curie temperature if
quantum effects would not exist) and $B$ is a constant independent
of temperature. The resulting fits are indicated by means of solid
lines in Fig.~\ref{f:fig3BZ}, and the coefficients of these fits
are given in Table~\ref{t:TBarrett}. In addition to the fact that
the quality of these fits is in overall rather good (thus,
confirming, as in KTaO$_3$ and SrTiO$_3$, the relevance of the
empirical Barret relation for describing quantum effects on
dielectricity of perovskites), four features are particularly
worth noticing related to these fits. First of all, $T_0$ is
(strongly) {\it negative} which confirms our theoretical findings
that BaZrO$_3$, unlike KTaO$_3$ and SrTiO$_3$, does {\it not}
exhibit any ferroelectric instability even in the classical
regime. Secondly, the T$_s$ deduced from the experimental data is
around 75-81\,K, which is located in the temperature region for
which PI-QMC predictions begin to significantly differ from CMC
results for dielectric as well as structural properties (see
Fig.~\ref{f:fig3BZ} and Fig.~\ref{f:AlatT}). In other words, our
simulations confirm the physical meaning usually associated with
$T_s$, that is the temperature below which quantum effects play a
non-negligible role on physical properties
~\cite{Barrett,AliLaurent,Ang}. Thirdly, and also unlike in
KTaO$_3$~\cite{Ang} and SrTiO$_3$~\cite{KleemannQBarret}, the B
parameter can {\it not} be neglected to get good fits in
BaZrO$_3$. This latter difference is due to the fact that the
low-temperature plateau is much larger in KTaO$_3$ (around
4000)~\cite{AliLaurent,Salce,SamaraBook} and SrTiO$_3$ (around
20,000)~\cite{KleemannQBarret,Viana} than in BaZrO$_3$ (around 55,
see Fig.~\ref{f:fig3BZ}), or equivalently that $T_0$ is positive
in KTaO$_3$~\cite{Ang,AliLaurent} and
SrTiO$_3$~\cite{KleemannQBarret} while being negative in
BaZrO$_3$. Finally, two fine structures, existing in the
experimental data, deviates from (and seem to be superimposed with
respect to) the Barrett fit. More precisely, one hump appears in
$\varepsilon_1$ around 50\,K at the lowest used frequency {\it
versus} 65\,K for the highest frequency, while a second hump shows
up around 15\,K for (more-or-less) any frequency. The magnitude of
this second hump decreases when decreasing frequency.  Note that
these relatively small humps do {\it not} appear within our CMC
simulations, and that the large fluctuations inherent to the
PI-QMC approach~\cite{Cuccoli} do not allow us to assert if these
humps are also predicted from our quantum simulations.

\subsection{IMAGINARY PART OF THE DIELECTRIC RESPONSE AND DIELECTRIC LOSS} \label{s:Loss}
\begin{figure}
\includegraphics[height=0.4\textheight]{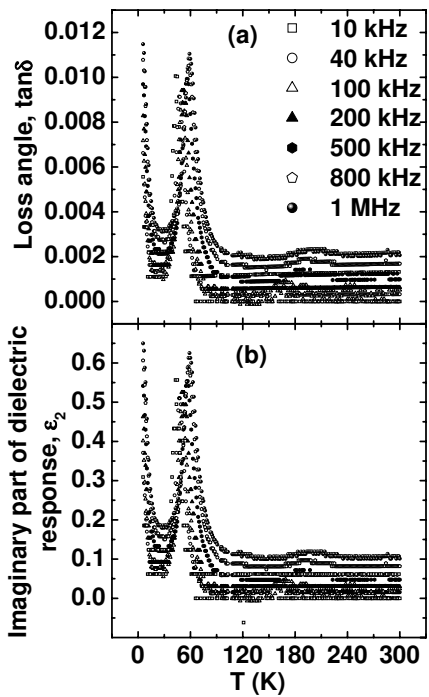}
\caption{Measured (a) loss angle, tan$\delta$, and (b) imaginary
part of the dielectric response, $\varepsilon_2$ at different
frequencies \emph{versus} temperature.}\label{f:fig4BZ}
\end{figure}

Fig~\ref{f:fig4BZ} displays the {\it imaginary} part of the
dielectric response, $\varepsilon_2$ and the loss angle (as given
by $\tan\delta=\varepsilon{_2}/ \varepsilon{_1}$) of BaZrO$_3$
versus temperature for frequencies ranging between 10\,kHz and
1\,MHz.
All the results displayed in Fig~\ref{f:fig4BZ} are from
measurements since simulating loss on an {\it ab-initio} level is
one of the most interesting and difficult challenges that remain
to be accomplished nowadays. One can first notice from
Fig~\ref{f:fig4BZ} that both $\varepsilon_2$ and  $\tan\delta$ are
very weak for the whole investigated frequency range, as also
observed in Ba(Zn$_{1/3}$Ta$_{2/3}$)O$_3$-BaZrO$_3$ solid
solutions~\cite{Davies}. BaZrO$_3$ can thus be a material of
choice to design high-Q compounds. Furthermore, our measurements
show the existence of two loss anomalies, namely a peak in
$\varepsilon_2$ (and $\tan\delta$) around 50-65\,K and a
continuously increasing $\varepsilon_2$ (and $\tan\delta$) when
decreasing temperature below 25\,K. These two anomalies are
correlated with the humps seen in $\varepsilon{_1}$  since they
occur at similar temperatures and since they behave in a  similar
fashion. For instance, both the peaks in $\varepsilon{_1}$ and
$\varepsilon{_2}$, occurring around 50\,K at low frequency, shift
to higher temperature when increasing the frequency. Similarly,
the anomalies observed in $\varepsilon{_1}$ and $\varepsilon{_2}$
at very low temperature get more pronounced at higher frequency.
Interestingly, the well-known Kramers-Kronig
relations~\cite{Scaife} imply that any peak/anomaly in
$\varepsilon{_1}$ should be accompanied with a peak/anomaly in
$\varepsilon{_2}$ (and vice-versa) at the same temperature. Such
correlation is thus indeed  satisfied in BaZrO$_3$, but
surprisingly, does not seem to hold for pure SrTiO$3$ and KTaO$3$.
More precisely, $\varepsilon{_2}$ has been found to exhibit a peak
in these two latter materials for a temperature around 30-50\,K,
but no corresponding peak has been seen in the real part of the
dielectric response. One possible reason for this lack of
observation  in pure KTaO$_3$ and pure SrTiO$_3$ may be due to the
fact  that these two materials (unlike BaZrO$_3$) have a  {\it
large} overall real part of the dielectric
response~\cite{Viana,AliLaurent} that ``washes out'' (i.e.,
prevents  the observation of) weak superimposed peaks.

Furthermore, the dielectric anomalies seen in BaZrO$_3$ and
reported in Figures~\ref{f:fig3BZ} and ~\ref{f:fig4BZ} are also
rather intriguing since our XRD and neutron scattering do not
detect any structural phase transition down to 2\,K. A discussion
about their possible causes is given in the next section.

\section{DISCUSSIONS}
\label{s:disc}

We believe that the enhancement in the real and imaginary parts of
the dielectric responses occurring at temperature around $\simeq$
15\,K is due to the activation of the oxygen octahedra rotation in
BaZrO$_3$. This belief is based on the fact that the minimum
energy associated with antiferrodistortive motions is predicted to
correspond to a temperature that is very close to the one at which
these enhancements begin to occur, namely 17\,K (see
Fig.~\ref{f:fig2BZ} and Sec.~\ref{s:Resstruc}). The zero-point
phonon vibrations may annihilate the {\it long-range order} of
such rotation down to 2\,K, which would explain our X-ray
diffraction and neutron scattering results. However, these
quantum-induced effects likely can not prevent the octahedra
rotation from occurring at a {\it local} scale (especially at very
low temperature), which would be consistent with the fact that
$\varepsilon_2$ increases when decreasing the temperature below
25\,K (see Fig~\ref{f:fig4BZ}).

Regarding the unusual dielectric features occurring around 50\,K,
it is important to realize that a frequency-dependent peak has
previously been reported for  $\tan\delta$ and $\varepsilon{_2}$
in several other perovskite systems (e.g., BaTiO$_3$:La,
SrTiO$_3$:La, SrTiO$_3$, SrTiO$_3$:Ca, K(Ta,Nb)O$_3$,
(Pb,La)TiO$_3$:Cu, KTaO$_3$) {\it near a similar temperature}
(see, e.g. Refs~\onlinecite{Bidault,Viana,samara,Salce} and
references therein). For instance, $\varepsilon{_2}$ peaks around
40\,K for low frequency in KTaO$_3$~\cite{samara,Salce}. It is
commonly believed that the reasons behind these peaks is the
existence of unavoidable impurity ions having a different valence
than the host atoms (see Refs.~\onlinecite{SamaraFerro04,Laguta}
and reference therein). In such a case, these peaks should become
more pronounced when intentionally doping the sample with
impurities, as consistent with the fact that adding up to 3\% of
Mn$^{2+}$ in KTaO$_3$ leads to  a noticeable peak in the real part
of the dielectric response around 40\,K \cite{SamaraFerro04}.
Observing and understanding the effects of doping on physical
properties of BaZrO$_3$ is thus of importance to confirm this
(general) possibility, but goes beyond the scope of the present
article. One particular previous study~\cite{Bidault} further
stipulates that it is a polaronic relaxation -- that is, a
coupling between the free charge carriers arising from the
impurity ions and the lattice properties of the host material --
that causes such weak dielectric anomalies. As done in Ref.
\onlinecite{Bidault}, such possibility can be checked by
extracting the temperature-dependency and frequency-dependency of
relaxation time via an analysis of loss dynamics using, e.g,
Cole-Cole formula loss~\cite{Scaife}. Such analysis, to be
accurate, requires the investigation of $\varepsilon{_1}$ and
$\varepsilon{_2}$ under a range of frequency that is much wider
than the one available for the present study.

\section{CONCLUSIONS}

In summary, we combined measurements with first-principles-based
techniques to investigate the low-temperature properties of
BaZrO$_3$. This system is found to be cubic and macroscopically
paraelectric down to 2\,K.  Unlike the ``better-studied''
KTaO$_3$~\cite{AliLaurent,SamaraBook} and
SrTiO$_3$~\cite{ZhongPRB96,Viana} incipient ferroelectrics, the
zero-point phonon vibrations do {\it not} suppress
ferroelectricity in BaZrO$_3$. In other words, this latter
material is also paraelectric in the classical regime. On the
other hand, quantum effects lead to the saturation of the cubic
lattice parameter below $\simeq$ 40\,K.

Despite having no long-range-order structural phase transition,
BaZrO$_3$ exhibits the following striking dielectric features: (1)
the $\varepsilon_1$ real part of the dielectric response saturates
in overall at low temperature (namely, below $\simeq$ 40\,K) to a
value $\simeq$ 55. Our PI-QMC simulations are in rather good
agreement with experimental data, and show that such saturation is
caused by zero-point phonon vibrations; (2) the temperature
behavior of $\varepsilon_1$ can be well fitted by the empirical
Barrett relation when allowing the $B$ parameter to differ from
zero in this fit; (3) two peaks or fine structures are observed in
$\varepsilon_1$ around 50-65\,K and 15\,K, respectively. The first
peak shifts to lower temperatures when decreasing the frequency,
while the second one occurs at around the same temperature but
decreases in magnitude when decreasing the frequency; (4) these
two fine structures are associated with anomalies in the imaginary
part of the dielectric response, $\varepsilon_2$, which peaks at
around 50-65\,K while suddenly and continuously increases when
decreasing temperature below 15\,K.

By comparing with previously reported data in other perovskites
and adopting some related interpretations, we propose that the
highest-temperature dielectric anomalies are caused by defects
like oxygen vacancies and/or unavoidable impurity ions such as
Fe$^{3+}$ --- which, e.g., are the source of free charge carriers,
that can interact with a soft lattice to create a polaronic state
responsible for the dielectric anomaly~\cite{Bidault}.
Furthermore, our (direct) first-principles calculations suggest
that the lowest-temperature dielectric anomalies may result from
local rotation of the oxygen octahedra.

In order to acquire a deeper knowledge of perovskites, it is
interesting to compare items (1-4) with corresponding features in
KTaO$_3$ and SrTiO$_3$. For instance,  $\varepsilon_1$ also
saturates  at low temperature in KTaO$_3$ and SrTiO$_3$, but with
a much higher value of the plateau (as a result of the
quantum-induced {\it suppression} of ferroelectricity in these
latter compounds). Such high value of the plateau explains why the
$B$ coefficient is generally omitted in the Barrett fit of
KTaO$_3$ and SrTiO$_3$. It is also highly plausible, as we believe
it, that such high $\varepsilon_1$  prevents the observation of
the hump that should be associated (according to the
Kramers-Kronig relation~\cite{Scaife}) with the peak of
$\varepsilon_2$ seen around 30-50\,K in KTaO$_3$~\cite{Bidault,Salce}
and SrTiO$_3$~\cite{Bidault,Viana}.

We hope that our work stimulates further investigations aimed at
checking our suggestions, in particular, and understanding dielectric
anomalies in perovskites, in general.
\label{s:concl}

\textbf{ACKNOWLEDGMENTS}

The authors would like to thank Jorge \'I\~niguez and David
Vanderbilt for providing the code for PI-QMC simulations. This
work is supported by ONR grants N 00014-01-1-0365 (CPD),
N00014-97-1-0048, N00014-04-1-0413 and 00014-01-1-0600, by NSF
grants DMR-0404335 and DMR-9983678 and by DOE grant
DE-FG02-05ER46188.


\begin{thebibliography}{99}
\bibitem{footnote} Dielectric constants at different
temperatures are typically calculated using the correlation
function method. However, for some temperatures and in order to
get better statistics and smoother curves, we also used the
difference between polarization under different electric field to
extract the dielectric constant. These two kinds of methods
provide similar results. For instance, in PI-QMC at T=75\,K
correlation function yield a value of 48.75 for dielectric
constant while on the other hand finite difference method lead to
49.48 when the electric field along either of x, y or z directions
is gradually increased from zero to 5$\times$10$^6$\,V/m.
\end{thebibliography}
\end{document}